\documentclass{llncs}
\usepackage{graphicx,wrapfig}
\usepackage{amsmath}

\DeclareMathOperator{\KS}{C}
\DeclareMathOperator{\poly}{poly}
\DeclareMathOperator{\polylog}{polylog}

\DeclareMathOperator{\prob}{Prob}

\DeclareMathOperator{\Ext}{Ext}
\DeclareMathOperator{\KExt}{KExt}
\DeclareMathOperator{\dep}{dep}
\DeclareMathOperator{\BT}{BT}
\DeclareMathOperator{\RBT}{RBT}

\newcommand{\U}{\mathcal{U}}
\newcommand{\V}{\mathcal{V}}
\newcommand{\eps}{\varepsilon}

\begin{document}

\title{Space-Bounded Kolmogorov Extractors\thanks{Supported by ANR Sycomore, NAFIT ANR-08-EMER-008-01 and RFBR~09-01-00709-a grants.}
}
\author{Daniil Musatov}
\institute{Moscow Institute for Physics and Technology and \\ Branch for Theoretical and Applied Research, Yandex LLC,
            \email{musatych@gmail.com}
}

\date{}
\maketitle

\begin{abstract}
An extractor is a function that receives some randomness and either ``improves'' it or produces ``new'' randomness. There are statistical and algorithmical specifications of this notion. We study an algorithmical one called Kolmogorov extractors and modify it to resource-bounded version of Kolmogorov complexity. Following Zimand we prove the existence of such objects with certain parameters. The utilized technique is ``naive'' derandomization: we replace random constructions employed by Zimand by pseudo-random ones obtained by Nisan-Wigderson generator.
\end{abstract}

\section{Introduction}
An extractor is a deterministic procedure that extracts randomness from weak random sources. Concerning finite strings there are two concepts of specifying this notion: statistical and algorithmical. The statistical one considers probability distributions on inputs and outputs of such procedure in terms of min-entropy. Loosely speaking, to extract randomness means to produce a distribution with higher min-entropy than any input has. This notion was invented by Nisan and Zuckerman in early 90s and was deeply examined by many researchers through the last two decades. An introduction to the field is presented by Shaltiel in~\cite{shaltiel}.

An algorithmic counterpart, i.e. the notion of Kolmogorov extractors, was invented in the last several years (see~\cite{FHPVW},~\cite{hitchcock} and~\cite{zimand10b}). Roughly speaking, a Kolmogorov extractor is a function that receives two strings with sufficiently large Kolmogorov complexity and sufficiently small dependency and outputs a sufficiently long string having complexity closer to its length than any input has. It was shown in~\cite{hitchcock} that there exists a deep connection between ordinary and Kolmogorov extractors. Namely, each ordinary extractor is a Kolmogorov extractor with a bit worse parameters and vice versa. As shown in~\cite{zimand09} and~\cite{zimand10}, there also exist \emph{strong} Kolmogorov extractors in a sense that the output is rather complex even being conditioned on any single input. 

The notion of Kolmogorov extractors may be naturally expanded to space-bounded complexity, it was done already in the original paper~\cite{FHPVW}. The existence results of that paper hold both for common and space-bounded Kolmogorov extractors. For the unbounded case these resuts were improved by Zimand in~\cite{zimand09} and~\cite{zimand10}. In this paper we convert Zimand's results to the space-bounded case and hence improve the respective results of Fortnow et al. Since Zimand's construction is not efficient, this conversion cannot be done straightforwardly. The technique we employ is the ``naive derandomization'' method introduced in~\cite{csr11-ar} and~\cite{csr11} and later used in~\cite{zimand11a} and~\cite{zimand11b}. Originally, Zimand have characterized Kolmogorov extractors by some combinatorial properties. The existence of an object with such properties was proven implicitly. We show that such an object may be found in the output of Nisan-Wigderson pseudo-random generator. That is, to find a required object one does not need to search through all possible objects but needs only to check all seeds of the generator. This crucially decreases the required space from exponential to polynomial.

The rest of the paper is organized as follows. In Sect.~\ref{pre} we give formal definitions of all involved objects and formulate necessary results. In Sect.~\ref{zimand} we give formal definitions for space-bounded Kolmogorov extractors, formulate our existence theorems, outline the proof idea and present detailed proofs.

\section{Preliminaries}\label{pre}
\subsection{Kolmogorov complexity}
Let $\V$ be a two-argument Turing machine. We refer to the first argument as to the ``program'' and to the second argument as to the ``argument''. (Plain) Kolmogorov complexity of a string $x$ conditioned on $y$ with respect to $\V$ is the length of a minimal $\V$-program $p$ that transforms $y$ to $x$, i.e. 
$$\KS_{\V}(x\mid y)=\min\{p\colon \V(p,y)=x\}$$
There exists an optimal machine $\U$ that gives the smallest complexity up to an additive term. Specifically, $\forall\V\exists c\forall x,y \KS_{\U}(x|y)<\KS_{\V}(x|y)+c$. We employ such a machine $\U$, drop the subscript and formulate all theorems up to a constant additive term. The unconditional complexity $\KS(x)$ is the complexity with empty condition $\KS(x\mid\varepsilon)$, or the length of a shortest program \textit{producing} $x$.

The next notion to be defined is resource-bounded Kolmogorov complexity. Loosely speaking, it is the length of a minimal program that transforms $y$ to $x$ efficiently. Formally, Kolmogorov complexity of a string $x$ conditioned on $y$ in time $t$ and space $s$ with respect to $\V$ is the length of a shortest program $p$ such that $\V(p,y)$ outputs $x$, works in $t$ steps and uses $s$ cells of memory. This complexity is denoted by $\KS^{t,s}_{\V}(x\mid y)$. Here the choice of $\V$ alters not only complexity, but also time and space bounds. Specifically, the following theorem holds:
\begin{theorem}
There exist a machine $\U$ such that for any machine $\V$ there exists a constant $c$ such that for all $x$, $y$, $s$ and $t$ it is true that $\KS^{s,t}_{\U}(x\mid y)\leq\KS^{cs,ct\log t}_{\V}(x\mid y)+c$.
\end{theorem}
In our paper we deal only with space bounds, so we drop the time-bound superscript in all notations.

\subsection{Extractors}
A $k$-weak random source of length $n$ is a random variable distributed on $\{0,1\}^n$ that has min-entropy not less than $k$, that is, any particular string occurs with probability not greater than $2^{-k}$. The statistical distance between two randomness distributions $\xi$ and $\eta$ on the same set $T$ is $\max_{S\subset T}|\xi(S)-\eta(S)|$.

Loosely speaking, a randomness extractor is a procedure that converts weak random sources to nearly uniform random sources. There are two common specifications of this notion: seeded extractor that gets a weak random source and a (small) truly random source and multi-source extractor that gets two weak random sources. The latter one is relevant to our paper, so we define it formally. A multi-source extractor with parameters ($n$, $m$, $k$, $\eps$) is a function $\Ext\colon\{0,1\}^n\times\{0,1\}^n\to\{0,1\}^m$ such that for any two independent $k$-weak random sources $x$ and $y$ the induced distribution $\Ext(x,y)$ is $\eps$-close to uniform. A multi-source extractor may be considered as a $2^n\times2^n$ table with each cell coloured in one of $2^m$ colours. It may be proven that the extractor property is equivalent to the following: for any set of colours (``palette'') $A\subset\{0,1\}^m$ in any rectangle $S_1\times S_2$, where $S_i\subset\{0,1\}^n$ and $|S_i|\ge 2^k$  the fraction of cells coloured in a colour from $A$ differs from $|A|/2^m$ by at most $\eps$. Yet another equivalent definition is the following: for any $Q\in[1,2^m]$ and any $S_1\times S_2$ the fraction of cells coloured in $Q$ most popular colours does not exceed $Q/2^m+\eps$.

\subsection{Balanced tables}
A balanced table is a combinatorial object considered by Zimand in papers~\cite{zimand08},~\cite{zimand09} and~\cite{zimand10}. In the last paper he has also introduced a slightly different object called rainbow balanced table. In some sense they are similar to multi-source extractors but use another notion of closeness of distributions. Here we present alternative definitions that seem more comprehensive though equivalent to original ones. 

A ($K$,$Q$)-balanced table is a function $\BT\colon\{0,1\}^n\times\{0,1\}^n\to\{0,1\}^m$ with the following property: in any rectangle $S_1\times S_2$, where $S_i\subset\{0,1\}^n$ and $|S_i|\ge K$ the fraction of cells coloured in $Q$ most popular colours is less than $2Q/2^m$. Contrasting to multi-source extractors, this property is less restrictive for big palettes ($|A|>\eps2^m$) and more restrictive for small ones ($|A|<\eps2^m$).

In~\cite{zimand10} Zimand introduces a variation of the above-defined object named rainbow balanced table. Here we present a bit different though equivalent definition of it. A ($K$,$Q$)-rainbow balanced table is a function $\RBT\colon\{0,1\}^n\times\{0,1\}^n\to\{0,1\}^m$ with the following property. Consider a rectangle $S_1\times S_2$ where $S_i\subset\{0,1\}^n$ and $|S_i|\ge K$. Let us mark in each row all cells coloured in one of $Q$ most popular colours of this row. Call the table row-rainbow-balanced if the fraction of marked cells in any rectangle is less than $2Q/2^m$. Then do the same with columns and call the table column-rainbow-balaced if the fraction of marked cells is again less than $2Q/2^m$. Finally, call the table rainbow-balanced if it is both row- and column-rainbow-balaced.

It was shown by the probabilistic argument that there exist balanced tables with $K\ge\sqrt{M}\poly(n)$ and $Q\ge\sqrt{M}\poly(n)$ and rainbow balanced tables with $K\ge M\poly(n)$ and any $Q$. 

\subsection{Nisan-Wigderson generators}
The Nisan-Wigderson pseudo-random generator is a deterministic polynomal-time function that generates $n$ pseudo-random bits from $\polylog(n)$ truly random bits. The output of such generator cannot be distinguished from truly random string by small circuits. Specifically, we exploit the following theorem from~\cite{NW97}:
\begin{theorem}\label{nw}
For any constant $d$ there exists a family of functions $G_n\colon \{0,1\}^k\to\{0,1\}^n$, where $k=O(\log^{2d+6} n)$, such that two properties hold:
\begin{description}
\item [Computability:] $G$ is computable in workspace $\poly(k)$ (that is, any particular bit of $G(x)$ may be found in this space);
\item [Indistinguishability:] For any family of circuits $C_n$ of size $\poly(n)$ and depth $d$ for any positive polynomial $p$ for all large enough $n$ it holds that:
$$\left|\prob_x\{C_n(G_n(x))=1\}-\prob_y\{C_n(y)=1\}\right|<\frac1{p(n)},$$
where $x$ is distributed uniformly in $\{0,1\}^k$ and $y$~--- in $\{0,1\}^n$.
\end{description}
\end{theorem}

By rescaling parameters we get the following
\begin{corollary}\label{nw-modified}
For any constant $d$ there exists a family of functions $G_n\colon \{0,1\}^k\to\{0,1\}^N$, where $k=\poly(n)$ and $N=2^{\poly(n)}$, such that two properties hold:
\begin{itemize}
\item $G$ is computable in polynomial workspace;
\item For any family of circuits $C_n$ of size $2^{\poly(n)}$ and depth $d$, for any constant $c$ and for all large enough $n$ it holds that:
$$\left|\prob_x\{C_n(G_n(x))=1\}-\prob_y\{C_n(y)=1\}\right|<2^{-cn}.$$
\end{itemize}
\end{corollary}

The last corollary implies the following basic principle:
\begin{lemma}\label{mainprinciple}
Let $\mathcal{C}_n$ be some set of combinatorial objects encoded by boolean strings of length $2^{O(n)}$. Let $\mathcal{P}$ be some property satisfied for fraction at least $\alpha$ of objects in $\mathcal{C}_n$ that can be tested by a family of circuits of size $2^{O(n)}$ and constant depth. Then for sufficiently large $n$ the property $\mathcal{P}$ is satisfied for fraction at least $\alpha/2$ of values of $G_n$, where $G_n$ is the function from the previous corollary.
\end{lemma}

\subsection{Constant-depth circuits for approximate counting}
It is well-known that constant-depth circuits cannot compute the majority function. All the more they cannot compute a general threshold function that equals $1$ if and only if the fraction of $1$'s in its input exceeds some threshold $\alpha$. Nevertheless, one can build such circuits that compute threshold functions approximately. Namely, the following theorem holds:
\begin{theorem}[\cite{ajtai}, \cite{viola}]\label{ajtai}
Let $\alpha\in(0,1)$. Then for any (constant) $\eps$ there exists a constant-depth and polynomial-size circuit $C$ such that $C(x)=0$ if the fraction of $1$'s in $x$ is less than $\alpha-\eps$ and $C(x)=1$ if the fraction of $1$'s in $x$ is greater than $\alpha+\eps$.
\end{theorem}
Note that nothing is promised if the fraction of $1$'s is between $\alpha-\eps$ and $\alpha+\eps$. So, the fact that $C(s)=0$ guarantees only that the fraction of $1$'s is at most $\alpha+\eps$, and $C(s)=1$~--- that it is at least $\alpha-\eps$.

\section{Main result}\label{zimand}

\subsection{Overview}
In this section we give all necessary definitions, observe existing results and formulate our theorem.

Let us formally define a Kolmogorov extractor. Dependency between $x$ and $y$ is defined as $\dep(x,y)=\KS(x)+\KS(y)-\KS(x,y)$. A computable function $\KExt\colon\{0,1\}^n\times\{0,1\}^n\to\{0,1\}^m$ is a ($k$,~$\delta$)-Kolmogorov extractor if for any $x$ and $y$ of length $n$ if $\KS(x)>k$, $\KS(y)>k$ and $\dep(x,y)<\delta$ then $\KS(\KExt(x,y))>m-\delta-O(\log n)$. Say that $\KExt$ is a strong ($k$,~$\delta$)-Kolmogorov extractor if, moreover, $\KS(\KExt(x,y)|x)>m-\delta-O(\log n)$ and $\KS(\KExt(x,y)|y)>m-\delta-O(\log n)$.

Zimand has proven that there exist Kolmogorov extractors with parameters close to optimal:
\begin{theorem}[\cite{zimand09}, \cite{zimand10}]
Let $k(n)$ and $\delta(n)$ be computable functions, such that $1<k(n)<n$ and $1<\delta(n)<k(n)-O(\log n)$. Then there exists a ($k(n)$,~$\delta(n)$)-Kolmogorov extractor for $m=2k(n)-O(\log n)$ and a strong ($k(n)$,~$\delta(n)$)-Kolmogorov extractor for $m=k(n)-O(\log n)$.
\end{theorem}

We rewrite the definitions and the theorem in the case of space-bounded complexity.  Since the difference $\KS^s(x)-\KS^s(x|y)$ is not monotone in $s$, we get rid of explicit usage of the term ``dependency''. Instead we say that a computable function $\KExt$ is a ($k$,~$\delta$)-Kolmogorov extractor with a space bound $s=s(n)$ if $\KExt$ is computable in space $O(s(n))$ and for some constant $\mu>1$ (not dependent on $x$, $y$, $k$, $s$, but possibly dependent on $\KExt$) if $\KS^s(x)>k$, $\KS^s(y)>k$, and $\KS^{\mu s}(x,y)>\KS^s(x)+\KS^s(y)-\delta$ then $\KS^{s}(\KExt(x,y))>m-\delta-O(\log n)$. If, moreover, $\KS^s(\KExt(x,y)|x)>m-\delta-O(\log n)$ and $\KS^s(\KExt(x,y)|y)>m-\delta-O(\log n)$ then the Kolmogorov extractor is strong.\footnote{Fortnow et al. do use the term ``dependency'' in~\cite{FHPVW} but define it for two distinct space bounds that correspond to $s$ and $\mu s$ in our definition.} We increase the space limit from $s$ to $\mu s$ in the definition of ``dependency'' since the space limit is determined up to a multiplicative constant dependent on the description method. We prove the following:

\begin{theorem}
There exists a polynomial $p(n)$ such that for any space-constructible function $s(n)>p(n)$ and any computable in space $s(n)$ functions $1<k(n)<n$ and $1<\delta(n)<k(n)-O(\log n)$ there exists a ($k(n)$,~$\delta(n)$)-Kolmogorov extractor with space bound $s(n)$ for $m=2k(n)-O(\log n)$ and a strong ($k(n)$,~$\delta(n)$)-Kolmogorov extractor for $m=k(n)-O(\log n)$.
\end{theorem}

\subsection{Proof idea}
In this section we retell Zimand's argument in space-bounded environment and emphasize what must be added to complete the proof.

We show that a certain balanced table is in fact a Kolmogorov extractor. The main idea is to obtain a contradiction between the hardness of $(x,y)$ and the simplicity of $T(x,y)$ by employing the balancing property. Let us come to more details. Let $d=\delta+c\log n$, where $c$ is a constant to be determined later. Take a ($2^k$, $2^{m-d}$)-balanced table $T\colon\{0,1\}^n\times\{0,1\}^n\to\{0,1\}^m$. Let $B_x=\{z\mid \KS^s(z) \le \KS^s(x)\}$, $B_y=\{z\mid \KS^s(z) \le \KS^s(y)\}$ and $A=\{w\mid \KS^s(w)<m-d\}$. Obviously, $(x,y)\in B_x\times B_y$. By the balancing property, the set $B_x\times B_y$ contains less than $2\cdot2^{\KS^s(x)+1}2^{\KS^s(y)+1}/2^d$ cells coloured in a colour from $A$, that is, in a colour with complexity less than $m-d$. (If necessary, expand $B_x$, $B_y$ and $A$ arbitrarily to obtain sets of required size and apply the balancing property). Hence, $(x,y)$ may be described by the table $T$, the sets $B_x$, $B_y$ and $A$, and the ordinal number of $(x,y)$ among cells in $B_x\times B_y$ coloured in a colour from $A$. By the balancing property the last number requires at most $\KS^s(x)+\KS^s(y)-d+3$ bits. The sets $B_x$, $B_y$ and $A$ are described completely by $n$, $\KS^s(x)$ and $\KS^s(y)$ and may be enumerated in space $O(s)$. If we obtain a table $T$ that has complexity $O(\log n)$ and may be evaluated in space $O(s)$ then we have $\KS^{O(s)}(x,y)<\KS^s(x)+\KS^s(y)-d+O(\log n)$ that contradicts the assumption $\KS^{\mu s}(x,y)>\KS^s(x)+\KS^s(y)-\delta$ for $\mu$ and $c$ large enough. So, the crucial missing component is a simple and space-efficiently computable balanced table. Without a space bound the existence of such table is proven by the probabilistic method and a simple table may be found by an exhaustive search: the first table in some canonical order has small complexity. Having a space bound added the construction should be derandomized.

To get the strong Kolmogorov extractor property we need to replace a balanced table by a rainbow balanced one with parameters ($2^k$,$Q=2^{m-d}$) where again $d=\delta+c\log n$. Let $A_v$ be the set $\{w\mid\KS^s(w|v)<m-d\}$. Obviously, $|A_v|<Q$. Call a string $v$ \emph{bad} if the fraction of cells in the row $B_x\times\{v\}$ coloured in a colour from $A_v$ is greater than $2\times 2^{-d}$. The fraction of cells coloured in one of $Q$ most popular colours is even greater, so there are less than $K$ bad rows, since otherwise the colouring of the rectangle $B_x\times\{\text{bad\ rows}\}$ contradicts the rainbow balancing property. If one could enumerate bad rows in space $s$ then all bad rows would have complexity at most $k$ and so $y$ would be a good row. If $T(x,y)\in A_y$ and $y$ is known then $x$ may be described by the table $T$, sets $B_x$ and $A_y$ and the ordinal number of $x$ among all cells in the row $y$ coloured in a colour from $A_y$. Since $y$ is good the last number requires less than $C^s(x)-d+1$ bits. The sets $B_x$ and $A_y$ are described completely by $n$ and $\KS^s(x)$ and may be enumerated in space $O(s)$. Finally, if $T$ has complexity $O(\log n)$ and may be evaluated in space $O(s)$ then we obtain a contradiction similar to the previous one. The whole argument may be repeated symmetrically for the complexity conditioned on $x$. As before, the crucial missing component is a simple and space-efficiently computable rainbow balanced table that also allows to enumerate space-efficiently ``bad'' rows and columns. We now turn to a high-level description of our derandomization method.

\subsection{Derandomization plan}
To derandomize the construction we use a ``naive'' idea of replacing a random construction by a pseudo-random one. This idea was originally presented in~\cite{csr11} and~\cite{csr11-ar}. The essence of the idea is to replace a brute-force search among all possible objects by a brute-force search in the output of the Nisan-Wigderson pseudo-random generator. Since the length of the seed is polylogarithmic in the size of the output the range of the search decreases crucially. To make the things work we should, firstly, prove that the necessary object exists among the output of the NW-generator and, secondly, prove that a good seed for the generator may be found efficiently. To prove the first thing we employ the basic principle~\ref{mainprinciple} that involves a constant-depth circuit to test the balancing property. The original balancing properties seem to be too hard to be tested by such circuits, so we weaken them. Specifically, for ordinary balanced tables we limit the balancing condition only to those rectangles and palettes being actually used in the proof. For rainbow balanced tables we go even further and directly specify the property used in the proof. Details follow in the next several subsections. In Sect.~\ref{refinement} we specify the weakening of the balancing property. In Sect.~\ref{existence} we prove that a modified balanced table exists in the output of the NW-generator. Next, in Sect.~\ref{searching} we show how to find a good seed in limited space. Finally, in Sect.~\ref{completion} we put all things together and finish the proof. We do all steps simultaneously for balanced tables (leading to Kolmogorov extractors) and rainbow balanced tables (leading to strong Kolmogorov extractors).

\subsection{The weakening of balancing conditions}\label{refinement}
Recall that a table is a function $T\colon\{0,1\}^n\times\{0,1\}^n\to\{0,1\}^m$. We refer to the first argument as to `column', to the second one as to `row' and to the value as to `colour'. Let $b$ be a positive number and $k<n$ and $q<m$ be some integers. Let there be a system $\mathcal{S}$ of pairs $(S,l)$ where $S$ is a subset of $\{0,1\}^n$ and $l\in[k,n]$ such that for any pair the set $S$ contains less than $2^l$ elements and the whole system contains $2^{\poly(n)}$ pairs. Let there also be a system $\mathcal{Q}$ of subsets of $\{0,1\}^m$ (i.e., palettes) such that any $Q\in\mathcal{Q}$ contains less than $2^q$ elements and the whole system contains $2^{\poly(n)}$ sets. Later we refer to sets in such systems as to \emph{relevant} ones. 
Say that a table $T$ is ($b$, $\mathcal{Q}$, $\mathcal{S}$)-balanced if for any $(S_1,l_1)$ and $(S_2,l_2)\in\mathcal{S}$ and for any $Q\in\mathcal{Q}$ the number of cells in $S_1\times S_2$ coloured in a colour from $Q$ is less than $2^{l_1+l_2+q-m+b}$. If the sizes of $S_1$ and $S_2$ are maximal, $Q$ is the set of $2^q$ most popular colours and $b=1$ then the bound matches the original one, i.e. the fraction of the popular colours is at most $2\times \frac{2^q}{2^m}$. The new parameter $b$ is introduced due to technical reasons and will be used in the next subsection. Take in mind that $b\approx1$.

The definition of rainbow balanced tables is modified in a more complicated way: we again fix a number $b$ and a system $\mathcal{S}$. Instead of a system $\mathcal{Q}$ of palettes we fix a system $\mathcal{R}$ of tuples of palettes. Here each palette contains less than $2^q$ elements, each tuple has length $2^l$ for some $l\in[k,n]$ and the whole system contains $2^{\poly(n)}$ tuples.
Take arbitrary sets $S_1$ and $S_2\in\mathcal{S}$ with corresponding $l_1$ and $l_2$ and a tuple $\vec Q=(Q_1,\dots,Q_{2^{l_2}})\in\mathcal{R}$. Then for each $i\in[1, |S_2|]$ mark those cells in $i$-th row of $S_1\times S_2$ coloured in a colour from $Q_i$. Say that a row is \emph{saturated} if it contains more than $2^{l_1+q-m+b}$ marked cells. We say that a table $T$ is ($b$, $\mathcal{R}$, $\mathcal{S}$)-row rainbow balanced if for any $S_1$, $S_2$ and $\vec Q$ the total number of marked cells in saturated rows is less than $2^{l_1+q-m+k+b}$. (In particular, there are less than $2^k$ saturated rows). We define a ($b$, $\mathcal{R}$, $\mathcal{S}$)-column rainbow balanced table similarly and say that a table is ($b$, $\mathcal{R}$, $\mathcal{S}$)-rainbow balanced if it is both ($b$, $\mathcal{R}$, $\mathcal{S}$)-row and ($b$, $\mathcal{R}$, $\mathcal{S}$)-column rainbow balanced.

One may easily see that for $b\ge1$ our modifications actually do strictly weaken both balancing properties, so a random table satisfies a modified property with even greater probability than the original one.

\subsection{Existence of balanced tables in the output of the NW-generator}\label{existence}
To prove that a modified (rainbow) balanced table exists in the output of the NW-generator we employ the basic principle~\ref{mainprinciple}. We present a constant-depth exponential circuit that tests the modified balancing property. By the basic principle, since the generator fools such circuits and a random table satisfies the modified balancing property with positive probability, the same holds for a pseudo-random one. A construction of such a circuit follows.

The modified balancing properties are tested rather straightforwardly. Since we do not need to build a uniform circuit we just hardwire the lists of relevant sets into the circuit. That is, we construct a circuit for a particular tuple $(S_1,S_2,l_1,l_2,Q)$ (or $(S_1,S_2,l_1,l_2,\vec Q)$), make $2^{\poly(n)}$ copies of this circuit for different tuples and take a conjunction. The construction of such a circuit follows.

It is rather easy to check whether a particular cell in $S_1\times S_2$ is marked. Indeed, we should check whether its colour coincides with one of those belonging to $Q$ (or $Q_i$) and take the disjunction of all results. The difficult point is to count the number of marked cells and to compare this number to $2^{l_1+l_2+q-m+b}$. (For rainbow balance we need to count marked cells in saturated rows and columns). This task cannot be solved exactly by constant-depth circuits but may be solved approximately. Fortunately, approximate solution is enough for our goal.

We employ a circuit existing by theorem~\ref{ajtai}. Specifically, for ordinary balanced tables this curcuit has $|S_1|\cdot|S_2|$ inputs, outputs 1 if there are less than $2^{l_1+l_2+q-m+1}$ ones among the inputs, outputs 0 if there are more than $2^{l_1+l_2+q-m+1.01}$ ones among the inputs and outputs any value otherwise. For rainbow balanced tables we use two such circuits sequentially. Firstly, we apply to every row a circuit with $|S_1|$ inputs that outputs 1 if there are less than $2^{l_1+q-m+1}$ ones among its inputs, outputs zero if there are more than $2^{l_1+q-m+1.01}$ ones and outputs any value otherwise. Secondly, we count the number of ones in rows that produce one on the previous stage. To this end, we take conjunctions of the output of the previous circuit with the inpits and apply a circuit with $|S_1|\cdot|S_2|$ inputs that returns 1 if it receives less than $2^{l_1+q-m+k+1}$ ones, returns 0 if it receives more than $2^{l_1+q-m+k+1.01}$ ones and returns any value otherwise. This construction is  repeated for columns and a conjunction of two values is taken.

The last subcircuit completes the description. Let us sketch the structure of the whole circuit once more. The input specifies colours of all $2^{2n}$ cells of the table. We also add constants for all possible colours from $\{0,1\}^m$. The full circuit consists of $2^{\poly(n)}$ identical blocks. Each block has two groups of inputs. The left group specifies colours of all cells in a particular rectangle $S_1\times S_2$. The right group specifies a palette $Q$ (or a set of palettes $\vec Q$). Different blocks are hardwired to different inputs and constants. For ordinary balanced tables each block consists of two levels. On the first level a simple equivalence circuit is applied to every pair of a colour from the left and a colour from the right. On the second level an approximate counting circuit is applied to the outputs of the first level. For rainbow balanced tables each block consists of five levels. The first level is the same. On the second level an approximate counting circuit is applied to the outputs of the first level separately for each row and for each column. On the third a conjunction of the outputs of the first two levels is taken separately for each row and for each column. On the fourth level another approximate counting circuit is applied to the outputs of the third level, separately for rows and columns. On the fifth level a conjunction of two results of the fourth level is taken. Finally, a conjunction is applied to outputs of all blocks. 

Clearly, these circuits have exponential size and constant depth. It is also clear that these circuits return one on (rainbow) balanced tables. Hence, they return one with positive probability on a random table. Hence, they return one with positive probability on a pseudo-random table produced by the NW-generator. If the first (resp., second) circuit returns one then the table is balanced (resp., rainbow balanced) for parameter $b=1.01$. In the next two subsections we specify the systems $\mathcal{S}$, $\mathcal{Q}$ and $\mathcal{R}$, show how to find a good seed for the generator and how to use a $1.01$-balanced and $1.01$-rainbow balanced tables to obtain the result.

\subsection{Searching for a good seed}\label{searching}
Until this point the construction was valid for any choice of the systems $\mathcal{S}$, $\mathcal{Q}$ and $\mathcal{R}$. The searching for a good seed cannot be performed for arbitrary systems, so now we specify them.

We take $\mathcal{S}$ to be the system of all pairs $(\{z\mid\KS^s(z)<l\},l)$ for $s=s(n)$ and $l\in[k,n]$. We take $\mathcal{Q}$ to be the system of all sets $\{z\mid\KS^s(z)<q\}$ for $s=s(n)$ and $q\in[1,m]$. Finally, we take $\mathcal{R}$ to be the system of all tuples $(\{z\mid\KS^s(z|v_1)<q\},\dots,\{z\mid\KS^s(z|v_{2^l})<q\})$ where $(\{v_1,\dots,v_{2^l}\},l)\in\mathcal{S}$. Clearly, the sizes of all sets, tuples and systems satisfy the requirements. Since $s(n)$ is space-constructible, the systems $\mathcal{S}$ and $\mathcal{Q}$ are enumerable in space $O(s)$. It means that there exists an $O(s)$-space algorithm that gets two numbers $i$ and $j$ and returns the $i$-th element of the $j$-th set in $\mathcal{S}$ (or $\mathcal{Q}$). If one of the numbers is out of range the algorithm returns an error message. A similar statement holds for $\mathcal{R}$, here the enumerating algorithm gets three numbers: the number of the tuple, the number of the set in the tuple and the number of the element in the set.

Since we care only about space, the problem of searching a good seed is equivalent to the problem of checking whether a seed is good. The crucial property of NW-generator that makes such a check possible in small space is that any bit of the output may be computed in polynomial space independently from all other bits. That is, one need not store the whole exponential output to check some local property. A detailed description of such a check follows.

After this point the constructions for ordinary and rainbow balanced tables are rather different. We start with the construction for ordinary tables. Firstly, let us notice that a seed is good if and only if it is good for any tuple $(S_1, S_2, l_1, l_2, Q)$. So, it is sufficient to sequentially check that a seed is good for any such tuple. A tuple is determined by the ordinal numbers of $(S_1,l_1)$ and $(S_2,l_2)$ in the enumeration of $\mathcal{S}$ and that of $Q$ in the enumeration of $\mathcal{Q}$. Having these numbers fixed, we sequentially generate colours of all cells in $S_1\times S_2$ and compare them to all colours in $Q$. Count the number of successive comparisons. Say that the tuple is good if this number is less than $2^{l_1+l_2+q-m+1.01}$. Since we are no more restricted to constant-depth circuits and may instead use any space-bounded computations the counting is made precisely. Since $\mathcal{S}$ and $\mathcal{Q}$ are enumerable in space $O(s)$, the generator uses another $O(s)$ portion of space, only space $O(n)$ is used for intermediate storage and $s$ is at least polynomial in $n$, the total space requirement sums up to $O(s)$. 

For rainbow balanced tables we sequentially check that a seed is good for any tuple $(S_1, S_2, l_1, l_2)$. The sets of palettes for row and column rainbow balanced properties are generated from $S_1$ and $S_2$ respectively: for each $S_1=\{x_1,\dots,x_L\}$ and $S_2=\{y_1,\dots, y_M\}$ we take $\vec{Q}_1=(\{z\mid\KS^s(z|y_1)<q\},\dots,\{z\mid\KS^s(z|y_M)<q\})$ and $\vec{Q}_2=(\{z\mid\KS^s(z|x_1)<q\},\dots,\{z\mid\KS^s(z|x_L)<q\})$. The subsequent check is performed by direct counting, as in the previous algorithm. The difference is that the counting proceeds in two stages: for any row (or column) the number of marked cells is counted, then it is determined whether the row (column) is saturated and the numbers for saturated rows (columns) are summed up. The used space is again $O(s)$. 

\subsection{Completion of the proof}\label{completion}
In this section we prove that the tables generated from seeds found by two algorithms in the previous subsection are indeed Kolmogorov and strong Kolmogorov extractors respectively.

Firstly we check the Kolmogorov extractor property. Fix the extractor parameters $k=k(n)$, $\delta=\delta(n)$ and $s=s(n)$. Let $d=\delta+c\log n$ where $c$ is a constant to be determined later and let $q=m-d$. Let $p$ be the seed found for parameters $k$ and $q$ in Sect.~\ref{searching}. We want to prove that $\KExt_p=NW(p)$ is a ($k$, $\delta$)-Kolmogorov extractor for space bound $s$. Firstly, note that it is computable in space $O(s)$: this space is enough both for finding $p$ and computing $\KExt_p(x,y)$. Secondly, let us prove the Kolmogorov extractor property. Take two strings $x$ and $y$ such that $\KS^s(x)>k$, $\KS^s(y)>k$ and $\KS^{\mu s}(x,y)>\KS^s(x)+\KS^s(y)-\delta$ where $\mu$ does not depend on $x$ or $y$ and will be determined later. To obtain a contradiction assume that $\KS^s(\KExt_p(x,y))<m-d$. Denote $l_1=\KS^s(x)$ and $l_2=\KS^s(y)$ and consider the sets $S_1=\{z\mid \KS^s(z)\le l_1\}$ and $S_2=\{z\mid \KS^s(z)\le l_2\}$. Each of them is relevant by construction. Denote $Q=\{z\mid \KS^s(z)<q\}$. This set is also relevant. By the choice of $p$ the rectangle $S_1\times S_2$ contains less than $2^{l_1+l_2+q-m+1.01}$ cells coloured in one of colours from $Q$. By the assumption and the definition of $S_1$ and $S_2$, the pair $(x,y)$ belongs to these cells. In this case $(x,y)$ may be described by $n$, $l_1$, $l_2$, $q$ and the ordinal number of $(x,y)$ among these cells. Indeed, having $n$ known we may find a good seed $p$; having $l_1$, $l_2$ and $q$ known we may search through cells $S_1\times S_2$ and check whether the current one has a colour from $Q$. The ordinal number specifies the needed cell. The total required space is $O(s)$. The total number of used bits is $l_1+l_2-d+O(\log n)$. So, we obtain $\KS^{O(s)}(x,y)<\KS^s(x)+\KS^s(y)-d+O(\log n)=\KS^s(x)+\KS^s(y)-\delta-c\log n+O(\log n)$ that contradicts the condition $\KS^{\mu s}>\KS^s(x)+\KS^s(y)-\delta$ for $\mu$ and $c$ taken large enough.

Next, we check the strong Kolmogorov extractor property. We again fix the extractor parameters $k$, $\delta$ and $s$. As before, let $d=\delta+c\log n$ and $q=m-d$. Let $p$ be again the seed found for parameters $k$ and $q$ (and the rainbow balancing property) in Sect.~\ref{searching}. We want to prove that $\KExt_p=NW(p)$ is a strong ($k$, $\delta$)-Kolmogorov extractor for space bound $s$. The computability in space $O(s)$ is again easily obtained. Prove the strong Kolmogorov extractor property. Take two strings $x$ and $y$ such that $\KS^s(x)=l_1>k$, $\KS^s(y)=l_2>k$ and $\KS^{\mu s}(x,y)>l_1+l_2-\delta$ where $\mu$ does not depend on $x$ or $y$ and will be determined later. To obtain a contradiction assume that $\KS^s(\KExt_p(x,y)|y)<m-d$. Thus, the cell $(x,y)$ is marked. Consider two cases: there are more than $2^{l_1-d+1.01}$ strings $z\in S_1$ such that $\KS^s(\KExt_p(z,y)|y)<m-d$ and there are not more than $2^{l_1-d+1.01}$ such strings. In the first case the row $y$ contains more than $2^{l_1+q-m+1.01}$ marked cells and thus is saturated. By the $1.01$-rainbow balancing property the total number of marked cells in saturated rows is less than $2^{l_1+q-m+k+1.01}<2^{l_1+l_2-\delta-c\log n}$. Then $(x,y)$ may be described in space $O(s)$ by its ordinal number among marked cells in saturated rows and numbers $n$, $l_1$, $l_2$, $q$. Thus for large enough $\mu$ and $c$ the complexity of $(x,y)$ is less than $l_1+l_2-\delta$ that contradicts the assumption. In the second case the pair $(x,y)$ may be described by a description of $y$ ($l_2$ bits), the ordinal number of $(x,y)$ among marked cells ($l_1-d+1.01$ bits) and numbers $n$, $l_1$, $l_2$, $q$ ($O(\log n)$ bits), totaling to $l_1+l_2-d+O(\log n)$ bits. The required space is $O(s)$, so we obtain a similar contadiction to the assumption that $\KS^{\mu s}(x,y)>l_1+l_2-\delta$ for large enough $\mu$ and $c$. This contradiction finishes the proof.

\begin{center}
        \textbf{Acknowledgments}
\end{center}
        \nopagebreak

I want to thank my colleagues and advisors Andrei Romashchenko, Alexander Shen and Nikolay Vereshchagin for stating the problem and many useful comments. I also want to thank two anonymous referees for careful reading and precise comments. I am grateful to participants of seminars in Moscow State University for their attention and thoughtfulness.

\end{document}